# How does Bike Absence Influence Mode Shifts Among Dockless Bike-Sharing Users? Evidence From Nanjing, China

**Hongjun Cui**
School of Civil and Transportation Engineering,
Hebei University of Technology, Tianjin, China, 300400
Email: cuihj1974@126.com

**Zhixiao Ren**
School of Civil and Transportation Engineering,
Hebei University of Technology, Tianjin, China, 300400
Email: 202011601018@hebut.edu.cn

**Xinwei Ma**
School of Civil and Transportation Engineering,
Hebei University of Technology, Tianjin, China, 300400
Email: xinweima@hebut.edu.cn

**Minqing Zhu (Corresponding Author)**
School of Civil and Transportation Engineering,
Hebei University of Technology, Tianjin, China, 300400
Email: june146664@gmail.com

## ABSTRACT

Dockless bike-sharing (DBS) users often encounter difficulties in finding available bikes at their preferred times and locations. This study examines the determinants of the users' mode shifts in the context of bike absence, using survey data from Nanjing, China. An integrated choice and latent variable based on multinomial logit was employed to investigate the impact of socio-demographic, trip characteristics, and psychological factors on travel mode choices. Mode choice models were estimated with seven mode alternatives, including bike-sharing related choices (waiting in place, picking up bikes on the way, and picking up bikes on a detour), bus, taxi, riding hailing, and walk. The findings show that under shared-bike unavailability, users prefer to pick up bikes on the way rather than take detours, with buses and walking as favored alternatives to shared bikes. Lower-educated users tend to wait in place, showing greater concern for waiting time compared to riding time. Lower-income users, commuters, and females prefer picking up bikes on the way, while non-commuters and males opt for detours. The insights gained in this study can provide ideas for solving the problems of demand estimation, parking area siting, and multi-modal synergies of bike sharing to enhance utilization and user satisfaction.

## INTRODUCTION

Bike-sharing services have proven effective in alleviating urban traffic congestion, parking challenges, the last-mile problem, energy consumption and excessive emissions (*1*, *2*). Currently, there are two major bike-sharing systems: Station-based bike-sharing and Dockless Bike-sharing (DBS) (*3*). Station-based bike-sharing operates through fixed service stations, with users required to pick up and return bikes at designated locations (*4*). In contrast, the DBS system integrates advanced Internet of Things (IoT) and GPS technologies with smartphones to allow users to park bikes virtually anywhere within the service area (*5*). As of September 2022, 1889 bike-sharing systems are operational worldwide, with an additional 222 systems under construction (*6*).

The problem of unavailability of bikes: the inability of users to find available shared bikes when and where they are needed remains a major obstacle to DBS development (*7–9*). (*10*) analyzed DBS data for Beijing and discovered that over 80% of districts experienced unmet demand. Furthermore, the 2019 UC Davis campus travel survey data indicated that approximately 70% of DBS users reported such issues (*11*). The problem of unavailability can cause a discrepancy between real demand and DBS transaction records, leading to an overestimation or underestimation of DBS service levels (*12*).Many scholars have utilized data-driven (*13*) or discrete event simulator methods (*14*) to extrapolate real demand, but the above methods are unable to quantify the factors that impact real demand. Subsequently, (*15*) used a Poisson model to reconstruct the migration process of users when there are no bicycles available for borrowing, and constructed the demand truncation and migration Poisson model (*15*) .However, there are still gaps in the understanding of users' travel choice behaviors when there are no bikes available for borrowing, which restricts the ability to optimize the system operation further.

In order to increase the usage frequency of bike sharing, much of the literature has designed various types of scenarios to understand the incentives/disincentives that affect bike sharing usage (weather, public transportation, distance, and price)(*16–18*). However, there is limited research on scenarios where users are in a situation where there are no available bikes. When DBS users cannot find available bikes at their preferred location, their choice of alternative travel modes is primarily influenced by two factors: the distribution of the shared bikes and the availability of alternative travel modes (*15*). Specifically, users may attempt to pick up bikes from nearby areas using app-based information (*15*), wait for bikes to become available as other users return them (*19*), or opt for alternative modes, such as buses, taxis, or walking, which may provide greater convenience (*20*, *21*). Unmet user demand not only diminishes satisfaction (*22*) but also reduces company profits (*23*, *24*). Consequently, understanding the factors influencing travel mode shifts among DBS users under shared-bike unavailability is crucial for enhancing service quality and minimizing user dissatisfaction.

This paper makes the following main contributions.

●It explores users' travel choice behavior in the context of shared-bike unavailability, incorporating three specific bike-sharing-related choices: waiting in place, picking up shared bikes on the way, and picking up shared bikes on a detour. It also examines four alternative transportation modes: bus, taxi, riding, hailing, and walking.

●It establishes an integrated choice model using latent variables based on multinomial logit (ICLV-MNL) to investigate how socio-demographic characteristics, trip attributes, and psychological factors (accessibility, tangibles, and social benefits) influence users' travel choice behavior under bike availability.

The rest of the paper is structured as follows. The subsequent section offers a literature review and identifies gaps in current research. Section 3 describes the survey methodology and data collection process, along with an overview of the model employed. Section 4 presents an analysis of model estimation results and proposes pertinent policy implications. Conclusions and limitations are discussed in Section 5.

## LITERATURE REVIEW

### Study on the unmet demand for the Dockless bike-sharing

The operation of the DBS system relies on accurate and comprehensive data. However, unmet demand due to demand interruptions can lead to incomplete and inaccurate data, thereby undermining the effectiveness of operational strategies. In recent years, researchers have utilized data-driven approaches (*10*, *14*) and statistical modeling (*15*) to address the impact of unmet demand on the accuracy of demand estimates, highlighting an increasing focus on this area of study. (*25*) introduced a data-driven approach that enhances demand forecast accuracy by using the average use of shared bikes in each period as a minimum estimate of true demand. Despite its benefits, this method fail to fully quantify the factors influencing true demand. Therefore, some researchers continue to seek theoretical estimations. (*26*) developed a user trip model that captures completed orders and accounts for churned and shifted demand based on the average station behavior and daily demand trends, thereby improving demand forecasting accuracy. When users face bike unavailability, they may travel to other pickup points or abandon their intended use, leaving these demands unrecorded in DBS system data. (*27*) incorporated the migration behavior of unmet demand into their rebalancing strategy in a way to optimize user satisfaction and vehicle utilization. (*28*) argue that the impact of unmet demand is not readily apparent in the net demand calculation process, as it is challenging to capture from transaction data. They defined the net demand process as a non-stationary stochastic process that determines the number of bicycles required to meet the service level requirement. (*14*) introduced a simulation-based approach to estimate the real demand, with simulation, bootstrapping, and subset selection used to uncover latent demand within usage data. (*15*) analyzed the user demand truncation and migration process by establishing a relationship between observed demand and actual demand. They also proposed a demand truncation and migration Poisson model, which demonstrated superior adaptability and accuracy. Subsequently, (*29*) constructed the Demand Recovery considering demand Truncation, Migration, and spatial Correlation model to more accurately estimate the real demand of the free-flow shared bicycle system by considering demand truncation, migration, and spatial correlation that provides important support for optimizing rebalancing strategies.

**Factors influencing the Dockless bike-sharing**

Research conducted to enhance DBS and identify factors that promote bicycling in bike-sharing development has consistently been a scholarly focus. Studies often explore socio-demographic characteristics, travel attributes, and attitudinal variables (*30–32*). Socio-demographic characteristics factors, such as age, gender, and income, have received significant attention in research on travel mode choice (*33–35*). (*36*) employed multivariate models to evaluate how DBS serves as a last-mile solution by linking public transit stations with final destinations. During surveys on transportation preference, travel attributes play a pivotal role in decision-making. For leisure and entertainment trips, individuals show a preference for DBS over commuting or educational travel purposes (*32*, *37*, *38*). (*24*) noted significant morning and evening usage peaks in DBS usage on workdays, indicating commuting as a primary use for weekday commuters. Studies have also examined the influence of walking distance to pick-up locations on DBS mode choice (*39*). (*40*)analyzed the selected station paths of bike-sharing users through a questionnaire, finding a strong preference for the nearest station. Due to the slower riding and walking speeds, travel distances and walking distances to pick up bikes negatively impact the attractiveness of bike-sharing (*41*). (*42*) utilized historical travel data from Paris and a structural demand model to estimate the impact of accessibility to docking stations on DBS usage, finding that users 300 meters away from the pick-up station (approximately 4 minutes) are 60% less likely to use the service compared to users closer to the station. Several studies have explored the impact of other modes of transportation (car sharing(*43*), public transportation(*16*, *44*), taxi(*45*)and e-bikes(*46*)) on the behavior of DBS users, including pricing schemes, combined use schemes with DBS, and incentive strategies. Additionally, many scholars have noted the importance of attitudinal variables in the choice of bike-sharing trips. Factors such as societal benefits, convenience, accessibility (*47*), satisfaction (*48*), and service quality (*49*) have been empirically proven to significantly influence the choice of DBS.

**Research gaps**

In summary, while studies have considered the impact of unmet demand on the estimation of the real demand, the specifics of users' transfer to neighboring areas due to shared-bike unavailability and their influencing factors remain unclear. Existing research has primarily focused on the relationship between DBS and other transportation modes, including interactions with public transit, and has analyzed the influence of socio-demographic attributes, travel behaviors, and psychological attitudes on shared bike usage.

However, little research has been conducted on how these factors influence users' travel mode shifts in the context of shared-bike unavailability, especially from a psychological perspective. A comprehensive understanding of these aspects is essential for accurate demand prediction, strategic placement of pick-up points, and the resolution of challenges related to multi-modal integration to enhance DBS utilization and user satisfaction.

The aim of the present study is to fill this gap by investigating three potential user behaviors in response to shared-bike unavailability: waiting in place, picking up bikes on the way, and picking up bikes on a detour. This study employs a hybrid choice model that integrates latent variables representing accessibility, tangibles, and the social benefits associated with bike-sharing. This study allows for the assessment of the impact of socio-demographic factors, travel attributes, and these latent variables on users' decision-making processes under shared-bike unavailability.

## METHODOLOGY

### Study area

As the capital of Jiangsu Province and the second-largest city in eastern China, Nanjing boasts a sophisticated public transportation network featuring a well-developed metro system that consists of 12 lines and 208 stations. The massive population and advanced multimodal transportation system create optimal conditions for the extensive development of shared bike systems. In 2013, the introduction of a shared bike system in Nanjing enabled renting and returning shared bikes based on an IC card (*50*). As of 2020, a total of 53.2 thousand bikes are available in the docked bike-sharing system, distributed across 1.5 thousand designated parking terminals. In late 2016, DBS services were introduced in Nanjing (*51*). By the end of December 2021, there were three main licensed bike-sharing operators in Nanjing, with a total of 305,000 registered bikes. The system experienced an average daily ridership of 320,000 throughout the year, with peak daily rides as high as 680,000 (*52*).

### Survey design and data collection

A survey based on a web-based questionnaire was conducted to reveal users' mode shifts under shared-bike unavailability and their attitudes towards shared bikes. Data collection was carried out via Wenjuanxing between March 1, 2023, and March 14, 2023. The survey gathered information on travel mode choices and various attribute variables. The data acquired for each participant included:

● Socio-demographic background: gender, age, income, education, occupation.

● Characteristics of trips: the purpose of the trip, frequency of shared bike use, duration of rides, frequency of shared-bike unavailability, and frequency of using the app to check the distribution of shared bikes.

● Participants' attitudes toward DBS: Perceptions of DBS service quality were assessed through factors such as accessibility (*53*), tangibles (*48*, *54*), and social benefits (*55*, *56*). These perceptions were then quantified on a Likert scale ranging from "strongly disagree" (1 point) to "strongly agree" (5 points).

● Travel mode choice: There are seven major travel modes for participants: waiting in place, picking up on the way, picking up on the detour, bus, taxi, riding hailing, and walking. Specifically, waiting in place means waiting for other users to complete their ride and return a shared bike, or waiting for a transportation truck to restock the station with bikes. Picking up on the way means refers to users looking for available shared bikes along their route. Picking up on a detour means users choosing to pick up bikes at a location not directly in the direction of the destination, thereby increasing the walking distance accordingly.

Table 1 presents the attribute levels designed in this study. Table 2 presents the variables included in the final models and the ways they were measured.

**Table 1 Levels of attributes in experiments**

| Attribute | Attribute levels |
|---|---|
| Weather | Bad weather (Rain), Good weather (Sun) |
| Travel distance | 500 m, 1000 m, 1500 m, 2000 m |
| Travel purpose | Work/Go school, Leisure and entertainment |
| Waiting time for shared bikes | 1 min, 3 min, 5 min |
| Picking up shared bikes on the way | 2 min, 4 min, 6 min |
| Picking up shared bikes on a detour | 2 min, 4 min, 6 min |
| Waiting time for bus | 5 min, 10 min, 15 min |
| Waiting time for taxi | 8 min, 15 min |
| Waiting time for ride hailing | 5 min, 10 min |
| Cost for shared bikes | Free, 1.5 CNY |

(1 USD = 7.18 CNY (accessed 21 July 2023).)

**Table 2 Explanatory Variables and Measurements**

| Variable | Measurement |
|---|---|
| Gender | 1 if gender is male, 0 if female |
| Age | 1 if age is “under 18” or “18–25” or “26–35”, 0 if “36–45” or “46–59” or “60 or above” |
| Education | 1 if educational level is “Middle school and below” or “High school”, 0 if “Bachelor and junior college” or “Master and above” |
| Income (CNY/month) | 1 if monthly income is “under ¥3000” or “¥3000-¥5000” or, 0 if “¥5000-¥10000” or “over ¥10,000” |
| Weather | 1 if weather is rainy, 0 if otherwise |
| Commute | 1 if trip purpose is commute (i.e. work/education), 0 if otherwise |
| Travel cost | In RMB |
| Travel time | In min |
| Access time | In min, walking time to stations of bike share/ waiting time of taxi or bus |

Using an orthogonal main effects design, 32 experimental scenarios were generated. Figure 1 illustrates scenarios of shared-bike unavailability. This study concentrates on short-distance trips within 2 km given their prevalence in DBS travel in China (*57*). A total of 583 participants participated in the survey. After the exclusion of incomplete questionnaires, 551 valid questionnaires were obtained, with a sample of 2204 travel choices.

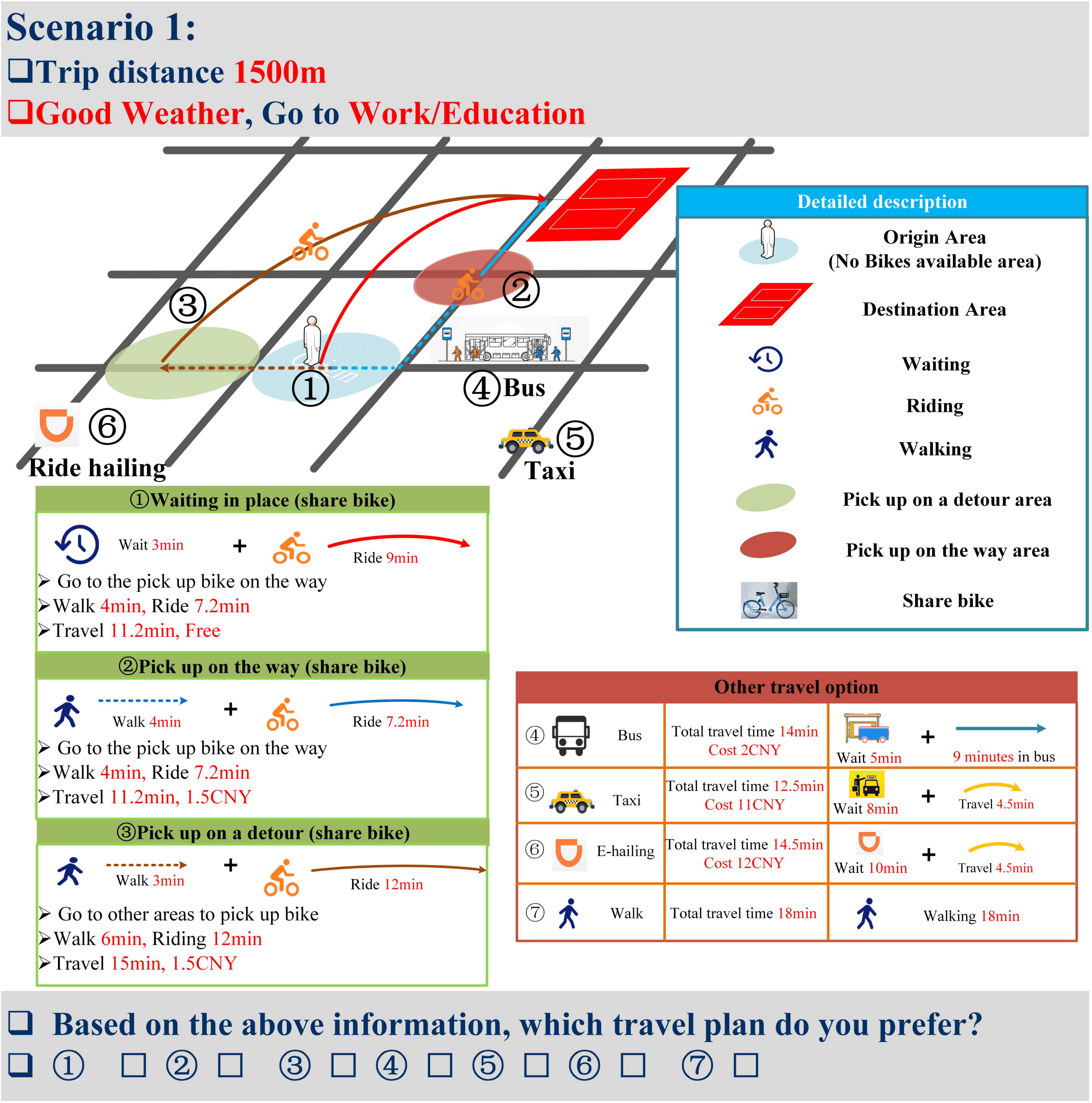

**Figure 1 Examples of shared-bike unavailability scenarios**

**Model structure**

MNL-based ICLV framework is employed to elucidate travel mode choice behavior under shared-bike unavailability. Figure 2 illustrates the structure of the ICLV-MNL model, which comprises two sub-models: the latent variable model and the discrete choice model (*58*, *59*). Measurement equations and structural equations, being integral components of model construction, work together to define the connections between latent variables and observed indicators, as well as the causal relationships among variables.

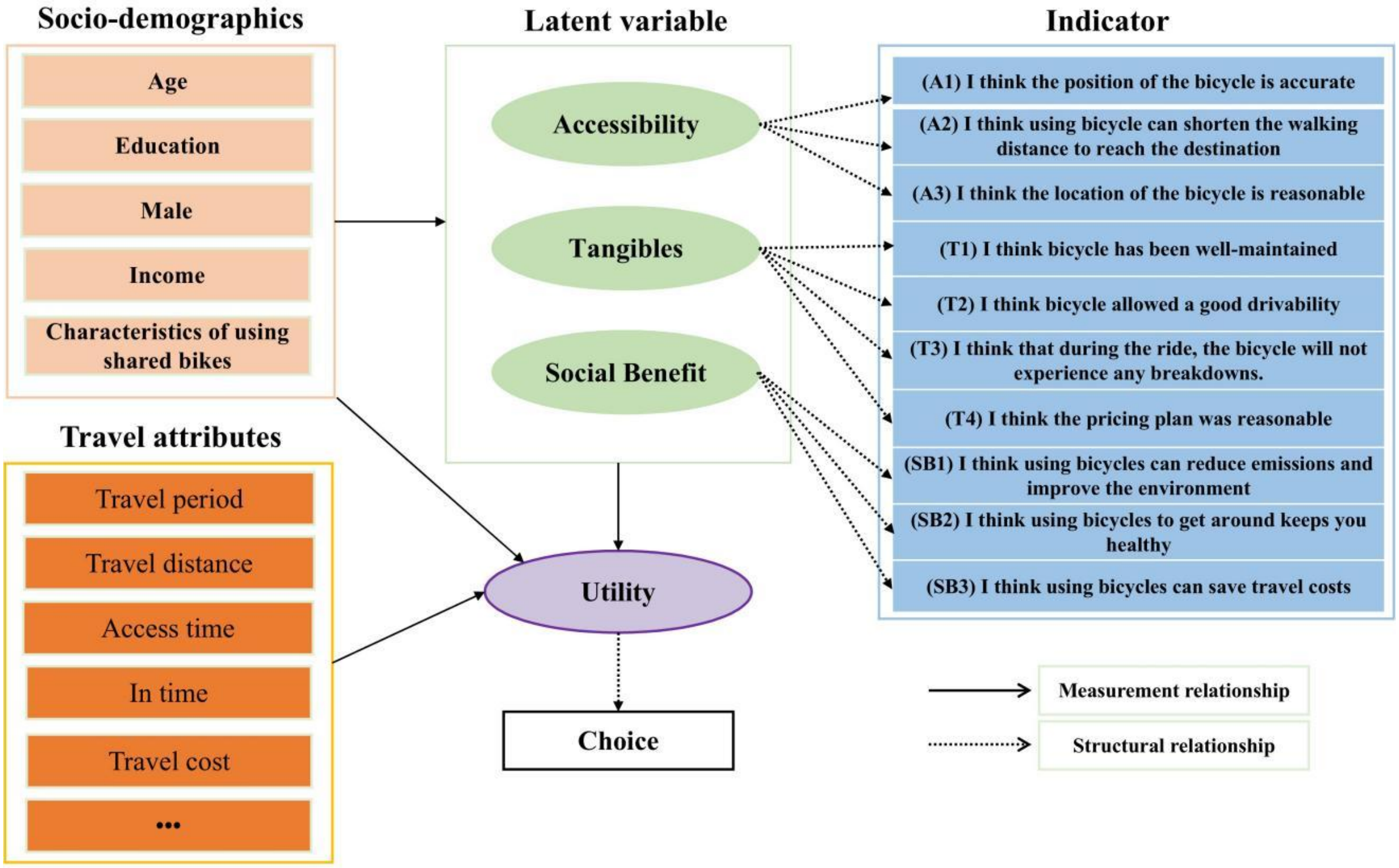

**Figure 2 Illustration of the ICLV-MNL model structure**

*Structural model*

In this model, three latent factors—accessibility, tangibles, and social benefits of DBS—are incorporated, each associated with a structural equation that establishes a connection between its value and the observed socio-demographic variables (*60*). We represent the latent variable $l$ of participant $n$ as $a_l^n$ (*16*) and express it in a linear formulation as **Equation 1**:

$$a_l^n = \zeta_l + \beta_l Z_n + \varphi_l^n, \varphi_l^n \sim N\left(0, \sigma_{\varphi_l}\right) \qquad (1)$$

Where $\beta_l$ and $\zeta_l$ are the coefficient vector and intercept of latent variable $l$ to be estimated, and $Z_n$ is the socio-demographics vector of participant $n$. $\varphi_l^n$ is the assumed stochastic error term of the latent variable $l$. It follows a standard normal distribution across participants, with a mean value of zero and a standard deviation of $\sigma_{\varphi_l}$.

*Measurement model*

Each response to a statement serves as an attitudinal indicator. The expression of its relationship with the corresponding latent variable requires a measurement equation. The specification of measurement models within the ICLV framework has been extensively discussed in existing research (*61*, *62*). This study employs an ordered probit model to clarify the observed indicator value, with the attitudinal statements rated on the 5-level response scale. Our choice to deviate away from simplistic linear approaches or deterministic methods, which might introduce biases, is justified. Thus, the indicator for the *s-th* attitudinal statement of participant $n$ with latent variable $k$ as an explanatory factor, can be formulated as follows:

$$I_{l,s}^{n*} = \gamma_{l,s} + \beta_{l,s}\alpha_l^n + \eta_{l,s}^n, \eta_{l,s}^n \sim N\left(0, \sigma_{\eta_{l,s}}\right) \qquad (2)$$

$$I_{l,s}^{n} = \begin{cases} v_{l,s}^{1}, I_{l,s}^{n*} < \tau_{l,s}^{1} \\ v_{l,s}^{2}, \tau_{l,s}^{1} \le I_{l,s}^{n*} < \tau_{l,s}^{2} \\ \vdots \\ v_{l,s}^{x}, \tau_{l,s}^{x-1} \le I_{l,s}^{n*} < \tau_{l,s}^{x} \\ \vdots \\ v_{l,s}^{X}, \tau_{l,s}^{X-1} \le I_{ls}^{n*} < \tau_{l,s}^{X} \end{cases} \quad (3)$$

Where $\beta_{l,s}$ and $\gamma_{l,s}$ are the coefcient and intercept of $\alpha_l^n$ to be estimated for the $s$-*th* attitudinal statement; $\eta_{l,s}^n$ is the assumed stochastic error term of $I_{l,s}^n$, which is normally distributed with a mean value of zero and standard deviation of $\sigma_{\eta_{l,s}}$; $I_{l,s}^{n*}$ is the response to the $s$-*th* attitudinal statement from participant $n$; $v_{l,s}^x$ is the $x$-*th* ordinal scale of the $s$-*th* attitudinal statement; and $\tau_{l,s}^{x-1}$ and $\tau_{l,s}^x$ are the lower and upper thresholds of $I_{l,s}^{n*}$.

$$\tau_{l,s}^{x} = \tau_{l,s}^{x-1} + \delta_l^{x-1} \quad (4)$$

The probability of participant $n$ responding $v_{l,s}^x$ to the *s-th* attitudinal statement can be expressed as:

$$P_s^n(I_{l,s}^n = v_{l,s}^x) = P\left(\tau_{l,s}^{x-1} \le I_{l,s}^{n*} \le \tau_{l,s}^{x}\right) \quad (5)$$

$$P_s^n(I_{l,s}^n = v_{l,s}^x) = \chi\left(\frac{\tau_{l,s}^{x} - \gamma_{l,s} - \beta_{l,s}\alpha_l^n}{\sigma_{\eta_{l,s}}}\right) - \chi\left(\frac{\tau_{l,s}^{x-1} - \gamma_{l,s} - \beta_{l,s}\alpha_l^n}{\sigma_{\eta_{l,s}}}\right) \quad (6)$$

Where $\chi(\bullet)$ represents the cumulative distribution function of the standardized normal distribution.

*Multinomial Logit model*

The discrete choice model pertains to a statistical approach used to examine the inclination of decision-makers to maximize utility through their choices. MNL model (*63*) emerges as a frequently employed statistical instrument for discerning the interplay among various factors in decision-making processes (*64*). For each participant $n$, there is a set of mutually exclusive alternatives $i$. $P_i$ denotes the probability of choosing alternative $i$.

$$P_i = \frac{\exp(U_i)}{\sum_{i=1}^{k} \exp(U_i)} \quad (7)$$

$$U_i = V_{n,i} + \varepsilon_{n,i} \quad (8)$$

Where $V_{n,i}$ is a systematic (observable) component of the utility function, $\varepsilon_{n,i}$ is a random disturbance (error term). In the ICLV model, which incorporates the effects of latent variables on mode utility (*16*), $V_{n,i}$ is represented in the following mathematical form:

$$V_{n,i} = \delta_{n.i} + \beta_i x_{n,i} + \lambda_i a_{n,i} \quad (9)$$

Where $\delta_{n.i}$ is an alternative (mode)-specific constant and $x_{n,i}$ is a vector of observed variables (e.g., socio-demographic and mode-specific variables) of participant $n$. $a_{n,i}$ is a vector of the latent variables of participant $n$, and $\beta_i$ and $\lambda_i$ are vectors of estimated mode-specific coefficients. Note that **Equation** 9 has an additional term $\lambda_i a_{n,i}$ , which incorporates the effects of the latent variables, unlike the conventional multinomial logit model.

**RESULTS**

Table 3 presents the socio-demographic characteristics and cycling habits of survey participants. The results show that 53.18% of participants were female, which is consistent with the known characteristics of DBS usage (*65*). The sample's largest age group, which makes up around 77% of the population, is 28 to 36 years old, indicating that bike-sharing users are often younger individuals. According to the survey, the participants are highly educated, with 94% holding at least a bachelor's degree. This corroborates the conclusions drawn by (*66*). Regarding usage frequency, most participants use the bike-sharing service 1-2 days (46%) and 3-4 days (29%) per week. Commuting emerges as the primary purpose for 59% of users, in line with prior studies (*67*). In terms of service engagement, over 24% of users carefully check bike locations before utilizing shared bikes, and over 90% have experienced difficulty finding DBS options, which is consistent with the findings of (*68*).

**Table 3 Sample characteristics**

| Question | Variable values | Percent (%) |
|---|---|---|
| Gender | Male | 46.82 |
| | Female | 53.18 |
| Age | Less than 18 | 0.18 |
| | 18-26 | 20.33 |
| | 26-36 | 57.35 |
| | 36-46 | 16.88 |
| | 46-60 | 5.26 |
| | More than 60 | 0 |
| Education | Middle school and below | 1.09 |
| | High school/former | 4.54 |
| | Bachelor and junior college | 85.48 |
| | Master or Doctor | 8.89 |
| Monthly Income (CNY) | Less than 3000 | 8.89 |
| | 3000-5000 | 15.25 |
| | 5000-1000 | 50.64 |
| | More than 10000 | 25.23 |
| Number of Own-car | 0 | 23.41 |
| | 1 | 69.15 |
| | 2 | 6.90 |
| | More than 2 | 0.54 |
| Number of bikes | 0 | 54.26 |
| | 1 | 42.29 |
| | 2 | 2.90 |
| | More than 2 | 0.54 |
| Number of electric bikes | 0 | 33.94 |
| | 1 | 52.63 |
| | 2 | 11.07 |
| | More than 2 | 2.36 |
| How often do you use shared bikes every week? | 1 day | 19.59 |
| | 2 day | 26.95 |
| | 3 day | 22.69 |
| | 4 day | 6.81 |
| | 5 day | 15.61 |
| | 6 day | 1.45 |
| | 7 day | 6.90 |
| Your average usage of shared bike time? | 0-5min | 7.44 |
| | 5-10 min | 27.40 |
| | 10-20 min | 42.10 |
| | 20-30 min | 18.33 |
| | More than 30 min | 4.72 |
| The purpose of using bike share in ordinary times? | Commute (Work/Go school) | 59.17 |
| | Non-commuting (Leisure and entertainment) | 40.83 |
| Have you ever experienced a situation where no bike is available? | Always | 1.27 |
| | Often | 16.15 |
| | Occasionally | 56.99 |
| | Once or twice | 16.15 |
| | Never | 9.44 |
| Do you view information (such as bike distribution) in the app when traveling using shared bikes? | Every time | 7.99 |
| | Often | 38.29 |
| | Occasionally | 29.58 |
| | Very seldom | 18.15 |
| | Never | 5.99 |

**Descriptive analysis**

Figure 3 illustrates the statistical results of travel mode choices based on individual preferences for different travel distances under shared-bike unavailability. Among the three DBS travel modes, picking up bikes on the way exhibits the highest percentage of travel choice preferences across different travel distances. This finding is consistent with the study by (*40*), likely influenced by the psychological inertia of the users' travel habits and their preference for picking up the bike without detouring. As travel distance increases, the popularity of picking up bikes on a detour decreases. There are two primary reasons: (1) Users may prefer more direct routes, and are unwilling to expend additional physical effort and time on detouring to pick up bikes as well as riding; (2) Users may lack trust in DBS apps and are concerned about failure to find available bikes (*68*). In addition, buses are likely to be the preferred solution for the majority of unmet demand, as the commuting expenses of buses and bike-sharing services are roughly comparable. It is worth noting that walking is the most preferred mode of travel for participants when the travel distance is 500 m.

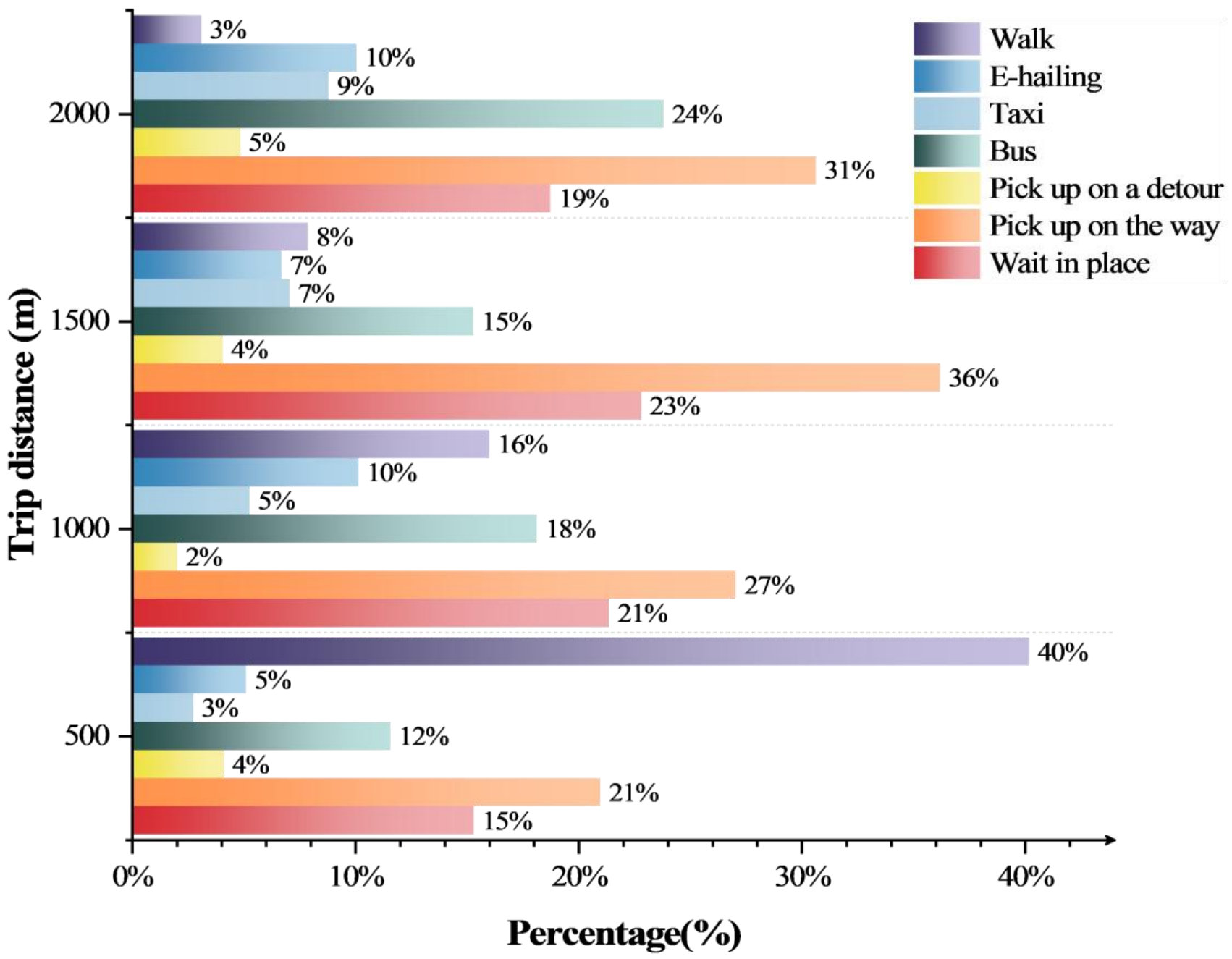

**Fig. 3 Travel choices for different travel distances**

Figure 4 illustrates the frequency of checking the distribution of shared bikes on apps. The analysis reveals that male participants prefer to use applications or applets to check this information before using the bike-sharing service compared to female participants. This indicates that males are more attentive to the location of shared bikes and have a greater awareness of the distribution of shared bikes near their location.

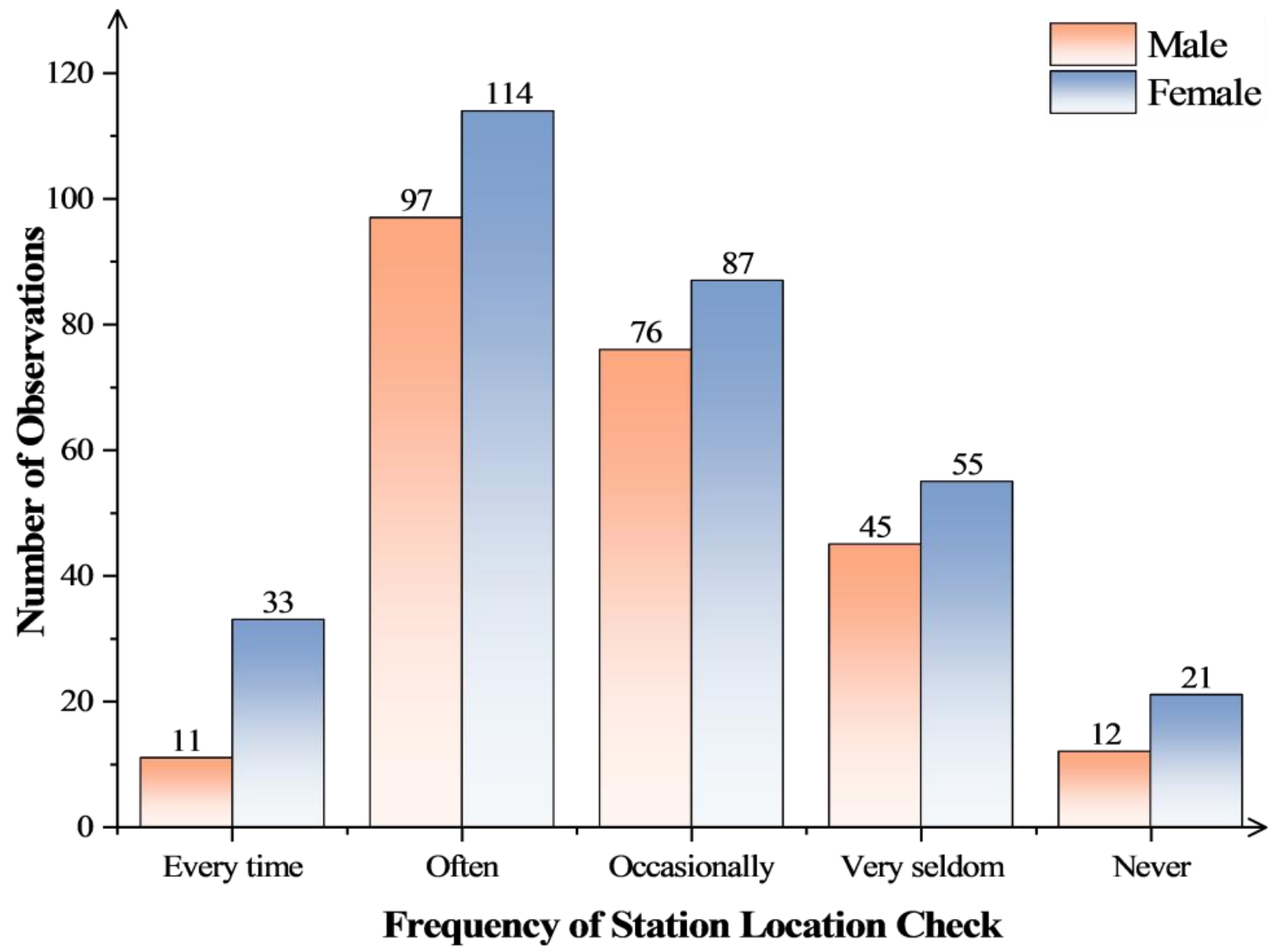

**Figure 4 The frequency of checking the distribution of shared bikes on apps**

Figure 5 presents statistics on the acceptable waiting and walking times for different travel distances. For the wait-in-place option, the percentage of users willing to wait for both 3 and 5 minutes increased, with a more pronounced rise for the 5-minute option (rising from 11% for 500 m to 29% for 2000 m), indicating that users are more tolerant of longer wait times for longer trips.

For the "Picking up on the way" option, the proportion of users willing to walk 4 minutes to pick up shared bikes increases with distance (from 15% at 500 m to 35% at 2000 m). However, the willingness to walk 6 minutes remains stable across distances. This indicates that the upper limit of acceptable walk time to pick up bikes under shared-bike unavailability is around 4 minutes (approximately 320 m), aligning with NACTO's findings NACTO, 2015, which suggest a 5-minute acceptable walking distance (1,000 feet).

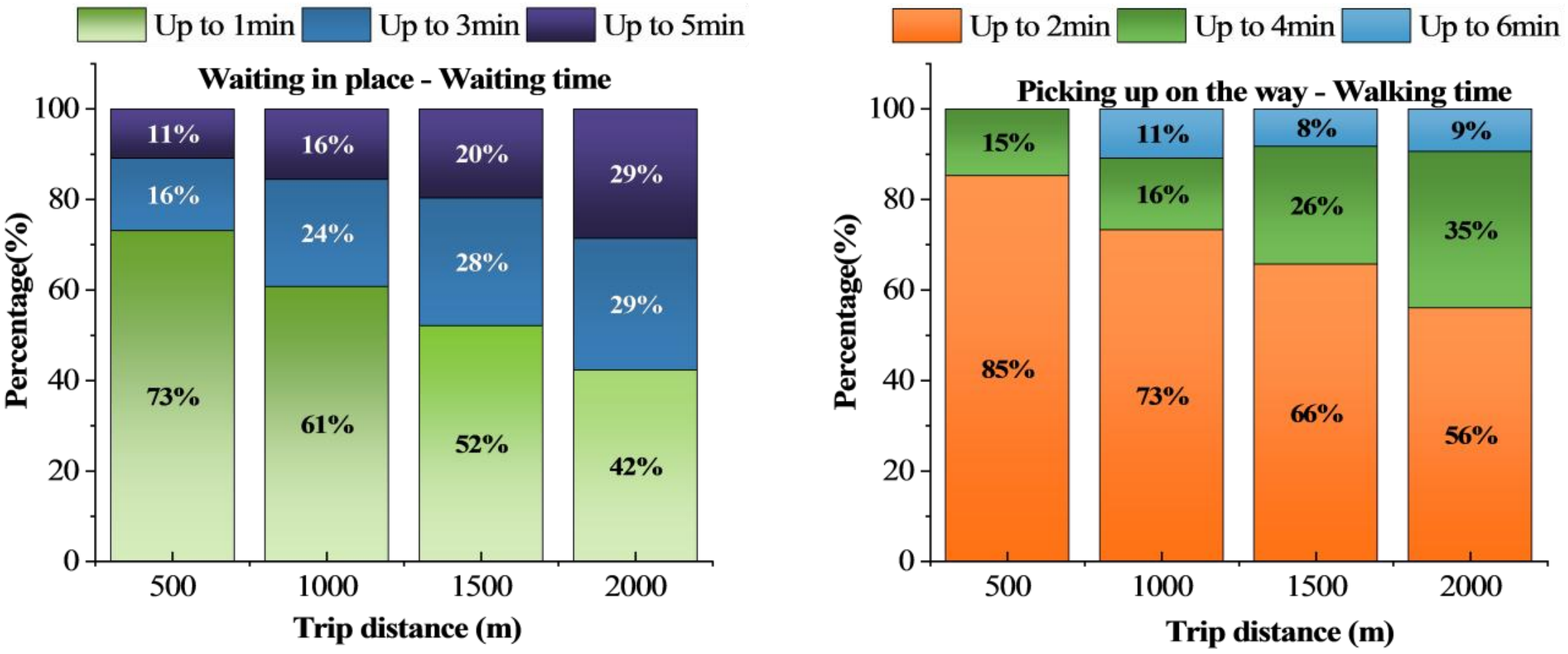

**Figure 5 Statistical results of waiting and walking time**

**Attitudinal statements analysis**

Table 4 shows the statistical distribution of participants' attitudes toward DBS, indicating that most remained positive towards all three aspects of bike-sharing. Among the indicators, participants rated the social benefits (SB) associated with DBS highly, particularly SB1 (4.48), SB2 (4.26) and SB3 (4.11). As the scale of bike-sharing expands, the maintenance and repair work becomes increasingly critical and complex. Approximately 23% of participants did not acknowledge the efforts in bike-sharing maintenance, possibly due to frequent encounters with broken bikes during use and delays in troubleshooting. Consequently, approximately 21% of participants remained negative toward the drivability of shared bikes. Inadequate maintenance of damaged bikes could significantly reduce user satisfaction and their willingness to use the service.

**Table 4 Statistical Distribution of Participants' Attitudes toward DBS**

| Variables and Statements | Distribution (%) | | | | | Likert scale | |
|---|---|---|---|---|---|---|---|
| | 1 | 2 | 3 | 4 | 5 | Mean | Std. |
| **Accessibility of DBS:** | | | | | | | |
| **A1:** I think the position of the bike is accurate. | 0.91 | 6.90 | 23.41 | 44.64 | 24.13 | 3.92 | 0.86 |
| **A2:** I think using bike can shorten the walking distance to reach the destination. | 1.09 | 4.72 | 16.51 | 52.45 | 25.23 | 3.78 | 0.91 |
| **A3:** I think the location of the bike is reasonable. | 1.27 | 8.35 | 23.95 | 46.10 | 20.33 | 3.69 | 0.93 |
| **Tangibles of DBS:** | | | | | | | |
| **T1:** I think bikes have been well-maintained. | 4.36 | 18.51 | 28.67 | 33.03 | 15.43 | 3.28 | 1.08 |
| **T2:** I think bikes allow good drivability. | 2.54 | 9.62 | 23.23 | 43.56 | 21.05 | 3.62 | 0.99 |
| **T3:** I think that during the ride, the bike will not experience any breakdowns. | 2.90 | 17.97 | 28.49 | 36.12 | 14.52 | 3.35 | 1.04 |
| **T4:** I think the pricing plan was reasonable. | 1.81 | 11.62 | 21.77 | 45.55 | 19.24 | 3.64 | 0.98 |
| **Social Benefit of DBS:** | | | | | | | |
| **SB1:** I think using bikes can reduce emissions and improve the environment. | 0.3 | 1.17 | 6.15 | 34.38 | 57.90 | 4.48 | 0.68 |
| **SB2:** I think using bikes to get around keeps you healthy. | 0.36 | 0.91 | 9.80 | 44.10 | 44.83 | 4.26 | 0.72 |
| **SB3:** I think using bikes can reduce congestion. | 0.544 | 4.17 | 15.06 | 41.74 | 38.48 | 4.11 | 0.88 |

The reliability and validity of a questionnaire are essential metrics for its measurement quality, critical for the accuracy of the model and the reliability of its outcomes. Cronbach's Alpha, introduced by Cronbach in 1951 (*70*), and the Kaiser-Meyer-Olkin (KMO) measure are employed to assess reliability and validity, respectively. Cronbach's coefficient, a standard method for internal consistency analysis, indicates strong consistency among latent variables, with all exceeding 0.7, indicating high reliability (see Table 5). Exploratory factor analysis (EFA) was conducted to identify significant latent variables influencing dependency indicators, revealing a 3-factor structure (Table 5).

The KMO test evaluated the data suitability for factor analysis and the strength of correlation between the variables (*71*). In this study, the KMO value derived from the questionnaire data was 0.804, signifying a good correlation among the three variables encompassing the 11 selected indicators. This suggests that the questionnaire demonstrates good structural validity, and the chosen indicators effectively capture the characteristics and attitudes of the participants.

**Table 5 Statistical results of responses to the attitudinal statements**

| **Variables** | | **Statements** | **Exploratory factor analysis results** | | | **Cronbach's alpha** |
|---|---|---|---|---|---|---|
| Accessibility of DBS | A1 | I think the position of the bike is accurate. | 0.801 | | | 0.750 |
| | A2 | I think using bike can shorten the walking distance to reach the destination. | 0.691 | | | |
| | A3 | I think the location of the bike is reasonable. | 0.649 | | | |
| Tangibles of DBS | T1 | I think bike has been well-maintained. | | 0.763 | | 0.813 |
| | T2 | I think bike allowed a good drivability. | | 0.702 | | |
| | T3 | I think that during the ride, the bike will not experience any breakdowns. | | 0.700 | | |
| | T4 | I think the pricing plan was reasonable. | | 0.616 | | |
| Social Benefit of DBS | SB1 | I think using bikes can reduce emissions and improve the environment. | | | 0.785 | 0.785 |
| | SB2 | I think using bikes to get around keeps you healthy. | | | 0.694 | |
| | SB3 | I think using bikes can reduce congestion. | | | 0.580 | |

**ICLV-MNL Models estimation results**

The maximum likelihood method is utilized to estimate unknown parameters within the ICLV-MNL model framework. In parallel, MNL models are executed for comparative analysis against the results obtained from the ICLV-MNL model. Estimation of the ICLV-MNL model is based on the Apollo R-programming package given its user-centric and robust choice modeling capabilities. To assess the goodness of fit of our proposed ICLV-MNL model, we employ Rho squared. A value of Rho squared greater than 0.2 suggests a relatively good fit. The adjusted Rho-squared values for both models surpass 0.2, indicating reasonable model specifications. Notably, the adjusted Rho squared value for the ICLV-MNL model, which incorporates latent variables, is 0.3602, surpassing that of the MNL model. These findings suggest that the elaborated ICLV-MNL model, incorporating latent attribute variables, better explains travelers' choice behavior. Table 6 presents the estimated results of the ICLV-MNL model. The BIC is used to balance the fit and complexity of the model and the model with lower BIC value is better. According to the estimation results it shows that the BIC of ICLV-MNL model (10284.56) is lower than the BIC of MNL model (12225.72).

**Table 6 Model estimation results**

| Variables | MNL | | ICLV-MNL | |
|---|---|---|---|---|
| | Coef. | t-stat | Coef. | t-stat |
| ASC –Waiting in place | -0.87*** | -2.61 | -0.73*** | -3.35 |
| ASC –Pick up on the way | -2.45*** | -4.93 | -2.31*** | -4.61 |
| ASC –Pick up on a detour | -6.11*** | -6.36 | -5.98*** | -8.21 |
| ASC –Taxi | -3.25*** | -4.92 | -3.27*** | -5.94 |
| ASC –Riding hailing | -8.76*** | -3.12 | -8.90*** | -6.16 |
| ASC –Walk | 2.34*** | 5.67 | 2.32*** | 9.70 |
| Education –Waiting in place | 0.89** | 2.20 | 0.85** | 2.12 |
| Income –Pick up on the way | 0.66** | 2.01 | 0.71** | 2.20 |
| Gender –Pick up on the way | 0.38*** | 3.80 | 0.21* | 1.75 |
| Gender –Pick up on a detour | -0.497*** | -2.38 | -0.59*** | -2.65 |
| Use bike frequency–Waiting in place | 0.40* | 1.73 | 0.17*** | 4.77 |
| Use bike frequency–Pick up on the way | 0.51*** | 2.29 | 0.09** | 2.52 |
| Use bike time –Waiting in place | 0.10 | 1.62 | 0.13** | 1.98 |
| Use bike time –Pick up on the way | 0.01 | 0.05 | 0.18*** | 2.76 |
| Use bike time –Pick up on a detour | 0.31 | 1.02 | 0.53*** | 4.74 |
| **Trip and mode attributes:** | | | | |
| Travel purpose –Pick up on the way | 0.44*** | 3.26 | 0.43*** | 5.16 |
| Travel purpose –Pick up on a detour | -0.38 | -1.62 | -0.39* | -1.68 |
| Weather –Waiting in place | -2.76*** | -14.76 | -2.77*** | -14.81 |
| Weather –Pick up on the way | -2.40*** | -14.72 | -2.41*** | -14.77 |
| Weather –Pick up on a detour | -1.38*** | -5.44 | -1.39*** | -5.48 |
| Weather –Taxi | 0.75*** | 3.09 | 0.76*** | 3.10 |
| Weather –Riding hailing | 0.99*** | 4.18 | 0.98*** | 4.17 |
| Weather –Walk | -1.39*** | -8.67 | -1.40*** | -8.72 |
| Waiting time –Waiting in place | -0.27*** | -7.92 | -0.94*** | -7.93 |
| Walking time –Pick up on the way | -0.32*** | -8.16 | -0.68*** | -8.24 |
| Waiting time –Taxi | -0.04* | -1.83 | -0.02* | -1.77 |
| Waiting time –Bus | -0.19*** | -7.23 | -0.08*** | -7.21 |
| Waiting time –Riding hailing | -0.15*** | -4.98 | -0.20*** | -5.02 |
| (In-time) Riding time –Waiting in place | 0.03 | 0.57 | 0.14*** | 2.63 |
| (In-time) Riding time –Pick up on the way | 0.11** | 2.01 | 0.11** | 2.05 |
| (In-time) Riding time –Pick up on a detour | 0.17** | 2.22 | 0.18** | 2.27 |
| (In-time) Time on Bus | 0.20*** | 3.24 | 0.15*** | 3.19 |
| (In-time) Time on Taxi | 0.47*** | 2.98 | 0.46*** | 2.94 |
| (In-time) Time on Walk | -0.15*** | -6.17 | -0.16*** | -6.24 |
| Travel cost –Waiting in place | -0.87* | -1.79 | -0.82* | -1.80 |
| Travel cost –Pick up on the way | -2.88*** | -6.56 | -2.85*** | -6.55 |
| Travel cost –Pick up on a detour | -3.41*** | -3.91 | -3.43*** | -3.93 |
| Travel cost –Bus | -0.54** | -2.24 | -0.24*** | -3.55 |
| Travel cost –Riding hailing | -0.63** | -2.17 | -1.84*** | -2.81 |
| **Latent variables:** | | | | |
| Availability of DBS | - | - | 0.08* | 1.69 |
| Tangibles of DBS | - | - | 0.25*** | 3.72 |
| Social Benefit of DBS | - | - | 0.02*** | 2.15 |
| **Model summary:** | | | | |
| Number of observations | 4408 | | 4408 | |
| Rho-squared | 0.2340 | | 0.3622 | |
| Adj.Rho-squared | 0.2203 | | 0.3602 | |
| BIC | 12225.72 | | 10284.56 | |

**Note:** *** Statistically significant at 0.01; ** Statistically significant at 0.05; * Statistically significant at 0.1.

*Travel attributes*

The estimation results indicate that an increase in travel costs reduces the utility of all transportation modes except taxis. This finding is consistent with (*6*), who determined that travelers prefer modes of transportation with lower travel costs. The access time coefficient for the "picking up bikes on the way" option is negatively correlated, aligning with (*57*), who found that longer walking times discourage the use of shared bikes. Additionally, since users are already under shared-bike unavailability, they may distrust the bike-sharing service and concern the same issue along the route to the station (*68*). Travelers are more likely to choose modes with shorter walk times or wait times (*16*). For the in-time coefficient, its increase enhances the utility of all travel modes except walking.

**Table 7 Marginal effects and elasticities of trip attributes on travel mode choice**

| Attribute | Waiting/Walking time | | In time | | Travel cost | |
|---|---|---|---|---|---|---|
| | Marginal Effect | Elasticity | Marginal Effect | Elasticity | Marginal Effect | Elasticity |
| Waiting in place | -0.147 | -0.757 | 0.022 | 0.113 | -0.128 | -0.661 |
| Pick up on the way | -0.138 | -0.488 | 0.022 | 0.079 | -0.578 | -2.045 |
| Pick up on a detour | - | - | 0.006 | 0.173 | -0.122 | -0.313 |
| Taxi | -0.001 | -0.019 | 0.025 | 0.434 | - | - |
| Bus | -0.011 | -0.066 | 0.021 | 0.125 | -0.034 | -0.199 |
| Riding hailing | -0.015 | -0.184 | - | - | - | - |
| Walk | - | - | -0.023 | -0.131 | -0.134 | -1.694 |

In order to avoid the misleading effect of directly comparing the coefficients of different scale variables, we further calculated the marginal effects and elasticity. Table 7 presents the travel cost in the option of pick up on the way and pick up on a detour has the greatest impact on users' decisions, followed by the time spent walking to pick up a bike, and the time spent riding in the absence of a bike to borrow appears to be unimportant. The results show that users are more sensitive to price changes when picking up on the way or on a detour, with price being the main driver of their decisions. Specifically, a 1% increase in travel cost leads to a 2% decrease in the probability of choosing the "pick up on the way" option and a 0.31% decrease for the "pick up on a detour" option. In contrast, walking time has a smaller effect: a 1% increase in walking time decreases the probability of choosing "pick up on the way" by 0.49%, while its effect on "pick up on a detour" is not significant. This is consistent with (*72*) who found that during short trips, users are more likely to choose bike-sharing services when alternative modes of transportation are more costly.This high sensitivity may also reflect the economic preferences of bike sharing users, i.e., when picking up on the way, or picking up on a detour, they tend to choose the low-cost option to optimize the overall cost of the trip.

The marginal effects and elasticity results for waiting time for the wait in place option are higher than those for travel cost and riding time, indicating that a reduction in user waiting time increases the probability of choosing the option more than a price adjustment. Specifically, a 1% increase in waiting time leads to a 0.76% decrease in the probability of choosing the wait in place option, and a 1% increase in ride time leads to a 0.11% decrease in the probability of choosing the waiting in place option. This is supported by (*73*), who found a tendency to perceive time spent waiting for bus as more onerous than time spent in the bus. In addition, waiting for alternative transportation modes such as bus (*74*) or subways (*75*) often involves uncertainty, which can increase anxiety in the context of shared-bike unavailability. Consequently, this highlights the importance of wait time minimization for greater user satisfaction and less perceived waiting-related burden.

The trip purpose coefficient exhibits a negative correlation with the "picking up bikes on the way" option, indicating a tendency among users to travel to stations on the shortest path to their destination to pick up shared bikes. Users' daily commute can be stress-inducing and negative sentiment-evoking (*76*). Walking to a shared bike station along the route to their destination could reduce perceived travel time and time-perceived anxiety. Conversely, the trip purpose coefficient has a positive relationship with picking up bikes on a detour, suggesting that when the trip's intent is non-commutative, such as for leisure or recreational activities, travel time constraints are less stringent, and users are more relaxed psychologically, which allow them to consider more leisurely and indirect routes.

In terms of the weather coefficients, the findings align intuitively with (*20*), who concluded that adverse weather discourages the choice of the three shared bike and walking modes while increasing the attractiveness of taxis and riding-hailing. Although bus usage also shows a positive correlation, it is normalized to the base when the model is specified.

*Socio-demographic and attitude variables*

The coefficient of education level is positively correlated with waiting in place, and the marginal effect suggests that users with low levels of education have a 13.3% higher probability of choosing to wait in place than users with high levels of education. Existing research has shown that people with high education levels exhibit a greater interest in bike sharing (*77*), as well as a greater likelihood of choosing a shared bike as their primary mode of transportation (*78*). However, under shared-bike availability, the reluctance of highly educated users to wait in place may be attributed to their focus on time management (*79*), leading them to avoid wasting time during the waiting process.

**Table 8 Marginal effects of socio-demographic attributes on travel mode choice**

| Attribute | Education<br>Waiting in place | Income<br>Pick up on the way | Gender<br>Pick up on the way | Gender<br>Pick up on a detour |
|---|---|---|---|---|
| Marginal Effect | 0.133 | 0.143 | 0.042 | -0.021 |

Marginal effects in Table 8 indicate that an increase in the user's income level has an inverse effect on the probability of choosing to pick up bike on the way. Specifically, low-income travelers (monthly income <CNY 5000) have a 14.3% higher probability of choosing to pick up on the way than higher-income travelers. Individuals from higher-income groups tend to place greater emphasis on the user experience and the perceived cost of services (*68*), whereas those from lower-income groups face increased financial pressure and find the pricing strategies of DBS schemes especially appealing.

There are notable disparities in travel mode choices among different groups, with gender differences being particularly pronounced (*80*).Specifically, the gender coefficient shows a weakly significant positive correlation with pick up on the way and a significant negative correlation with pick up on a detour. According to the marginal effect results, females are 4.2% more likely than males to pick up on the way, while males are 2.1% more likely than females to pick up on a detour.The data in Section 4.1 of this paper shows that female users pay more attention to the location of shared bikes than male users when using shared bikes for traveling, especially in the high-frequency attention level (e.g., "Often" and "Occasionally"), the proportion of female users is 53.3% and 40.6% respectively, compared with 45.3% and 35.5% for male users. This indicates that female users have a higher demand for information about bike locations and stations during trips, and are more likely to choose bike stations on familiar paths (picking up bikes on the way). Additionally, previous studies suggest that females are more likely to make trip chains on their commutes (*81*), which makes them reluctant to deviate from their established routes.

The results indicate that three latent variables in the ICLV-MNL model are positively correlated with the three modes of bike sharing. These findings are consistent with the research by (*48*, *55*, *56*), suggesting that system availability and tangible aspects positively impact bike-sharing choices shared-bike unavailability, thereby increasing users' willingness to use DBS. First, the availability of the DBS reflects users' perceptions of service accessibility. When users perceive that shared bikes are readily available, they are more likely to choose to use them. Second, the tangibility of the DBS relates to users' perceptions of the quality, maintenance status, and comfort of using the bikes. These tangible characteristics enhance users' satisfaction and trust, thereby positively influencing their choices. Finally, the social benefits of the DBS represent users' perceptions of the environmental and social advantages of using shared bikes, such as alleviating urban traffic congestion and reducing greenhouse gas emissions. This latent variable reflects users' environmental awareness and sense of social responsibility, which motivates them to choose more sustainable modes of transportation. Therefore, to promote the use of shared bikes among users, operators should improve the quality of their services to ensure high availability and high-quality physical features of the bikes. In addition, by publicizing the social and environmental benefits of shared bicycles and enhancing users' environmental awareness and sense of social responsibility, operators can further increase users' willingness to use them..

**Policy implication**

In the context of shared-bike unavailability, understanding travelers' mode choices can significantly increase the use of shared bikes. Based on the assessment results from the ICLV-MNL model and empirical data, we offer the following recommendations for bike-sharing system operators.

Statistics on shared bike travel mode selection confirm that users typically prefer stations closest to their intended destinations. To enhance customer satisfaction and the efficiency of bike-sharing systems, stations should be located along popular routes and near intersections. Data indicate that approximately 24% of surveyed users do not use mobile apps or applets to check the locations or availability of shared bikes before use.

To improve the perception of bike locations among users, information boards can be set up near shared bike stations or intersections to display information such as the arrival time of transfer trucks and the number of bikes to be returned soon. The refresh interval for information boards should optimally be set within 3 minutes. Considering habitual usage patterns of bike-sharing software, operators could engage users with check-in challenges or app-based rewards for viewing information, thereby boosting interaction with the software.

Model estimates suggest that a reduction in user waiting time decreases churn more effectively than a travel cost cut. To achieve this, bike-sharing operators can set up high-quality service facilities around subway or bus stations to minimize users' perceived waiting time. They might consider the establishment of waiting amenities such as kiosks, seating arrangements, and shelters near bike parking stations, or the integration with existing facilities at bus or subway stations to enhance user satisfaction and the quality of DBS service. Additionally, as trip distances increase, users are more likely to choose buses, taxis, and ride-hailing services due to their shorter arrival times. To address this, operators can analyze transaction data to implement differentiated incentive policies for regions with high demand for long-distance travel and increase service attractiveness, as well as regional coverage. Concurrently, operators can collaborate with the government to launch integrated monthly passes or discount packages that combine public transportation, metro and shared bikes. The aim is to enhance users' travel efficiency and experience.

Statistical results of latent variables indicate dissatisfaction among users with the maintenance of shared bikes, which adversely impacts the riding experience. Irregular service of damaged bikes can trigger a knock-on effect of neglect, known as the broken window effect, leading to a poor service experience. To boost satisfaction with shared bikes, operators should enhance their maintenance efforts.

However, there is insufficient information that guides users to participate in bike repair, with most users not well-informed about the repair process and the incentive program (*82*). Therefore, we suggest that operators start by strengthening the publicity and guidance for participation in bike repair. They can use apps and advertisement spots at bus and subway stations to promote the bike repair process and introduce rewards for participation in bike repair.

## CONCLUSIONS

The DBS systems frequently face challenges related to uneven spatial-temporal distribution, resulting in users' failure to find available bikes at preferred times and locations. Understanding the behavioral factors influencing users' travel choices when faced with unmet demand can enhance service quality and reduce user attrition. However, there has been limited research on the behavioral characteristics of users' mode shifts under shared-bike unavailability and the influence of user attitudes on these shifts.

To address this gap, this study conducted an online questionnaire survey in Nanjing, China, and proposed three alternatives for bike-sharing travel: waiting in place, picking up bikes on the way, and picking up bikes on a detour. The ICLV-MNL model was utilized to analyze the causal relationships between explanatory variables and travelers' attitudes, incorporating attitudes as variables in the choice model. The statistical analysis revealed that the highest percentage of users preferred picking up bikes on the way across different trip distances, while the detour option was the least favored, indicating a preference for walking to parking locations near their destinations. The ICLV-MNL model results showed that participants with a travel purpose of commuting/schooling, low-income individuals, and females are more inclined to pick up bikes on the way, while leisure travelers and males are more likely to choose a detour. Furthermore, shorter walking distances and wait times significantly promoted the continuous use of shared bikes. Adverse weather conditions were identified as major barriers to the sustained use of shared bikes. Additionally, alternative transportation modes such as buses, taxis, and ride-hailing service became more attractive as the travel distance increased. Our findings highlighted that latent variables associated with DBS positively impact user attrition reduction. However, dissatisfaction with the maintenance of shared bikes in Nanjing gave rise to unsatisfactory riding experiences, necessitating a reevaluation and enhancement of the maintenance programs by operators (*48*). In the planning and design of bike-sharing stations, operators should aim to ensure that the station distances are within a maximum acceptable walking time of approximately 4 minutes (around 320 meters) to meet the walking time preference of the majority of users, thereby improving system convenience and user satisfaction.

Despite valuable insights yielded by this study, it also has limitations that warrant further investigation. First, it focuses on users' mode shifts upon their arrival at a bike-sharing station without available bikes; it considers the access times of the three DBS types but neglects the walking time before reaching the station. Future research should explore the impact of walking time on their travel choices, as variations in walking time may affect users' tolerance and lead to different conclusions. Second, despite considerations given to factors such as the accessibility, tangibles, and social benefits of DBS, the psychological factors influencing users under shared-bike unavailability, such as service quality perceptions, require further examination. Although the direct use of latent variables provides significant interpretative advantages, we acknowledge that incorporating interaction terms between latent variables and observable characteristics could offer further insights. Such interactions may reveal how specific individual traits amplify or attenuate the impact of latent constructs on choice behavior. Future studies should investigate users' travel choices and walking directions under shared-bike unavailability, estimate the real demand for shared bikes, and develop rebalancing strategies based on user incentives.

## ACKNOWLEDGMENTS

This work was supported by the [National Natural Science Foundation of China] under Grant [numbers 52172304 and 52372302 and 52172304 ]; [Social Science Foundation of Hebei Province] under Grant [number HB22YJ040].

**AUTHOR CONTRIBUTIONS**

The authors confirm contribution to the paper as follows: study conception and design: Hongjun Cui, Zhixiao Ren; data collection: Zhixiao Ren; analysis and interpretation of results: Minqing Zhu, Xinwei Ma; draft manuscript preparation: Minqing Zhu, Zhixiao Ren. All authors reviewed the results and approved the final version of the manuscript.

**CONFLICT OF INTEREST**

The authors declared no potential conflicts of interest with respect to the research, authorship, and/or publication of this article.